# Energy landscape of ubiquitin modulated by periodic forces: Asymmetric protein stability and shifts in unfolding pathways

P. Szymczak* and Harald Janovjak[‡]

* Institute of Theoretical Physics, Warsaw University, Hoża 69, 00-681 Warsaw, Poland,
email: Piotr.Szymczak@fuw.edu.pl

[‡] Department of Molecular and Cell Biology, University of California Berkeley,
279 Life Sciences Addition, Berkeley, CA 94720, USA

## Abstract

Biological forces govern essential cellular and molecular processes in all living organisms. Many cellular forces, e.g. those generated in cyclic conformational changes of biological machines, have repetitive components. However, little is known about how proteins process repetitive mechanical stresses. To obtain first insights into dynamic protein mechanics, we probed the mechanical stability of single and multimeric ubiquitins perturbed by periodic forces. Using coarse-grained molecular dynamics simulations, we were able to model repetitive forces with periods about two orders of magnitude longer than the relaxation time of folded ubiquitins. We found that even a small periodic force weakened the protein and shifted its unfolding pathways in a frequency- and amplitude-dependent manner. Our results also showed that the dynamic response of even a small protein can be complex with transient refolding of secondary structures and an increasing importance of local interactions in asymmetric protein stability. These observations were qualitatively and quantitatively explained using an energy landscape model and discussed in the light of dynamic single-molecule measurements and physiological forces. We believe that our approach and results provide first steps towards a framework to better understand dynamic protein biomechanics and biological force generation.

# Introduction

Mechanical forces drive diverse biological processes such as cell adhesion (1, 2), cellular differentiation (3), flagellar beating (4) and hearing (5). On molecular scales, forces control the synthesis, degradation and import of proteins (6, 7), packaging and replication of nucleic acids (8, 9), cytoskeleton organization (10, 11), asymmetric spindle positioning (12) and diverse forms of mechano-sensitive signaling (13). Direct measurements of the nano-mechanics of biomolecules, obtained using force probe techniques (14) or molecular dynamics simulations (15), have been revolutionizing the way we perceive biomechanical and biochemical processes (16). In cellular contexts forces are often generated by biological machines, which transform chemical energy into directed mechanical motions *via* repeated conformational changes (9, 10, 17). As a direct consequence of their working cycles, biological machines are repetitive force generators and it is thus believed that periodic forces are experienced by biomolecules in a broad range of physiological contexts (3, 18-24). For instance, optical trapping experiments directly revealed that the forces generated by the motor protein dynein are oscillatory (22). Similarly, it has been postulated that periodic forces are utilized by proteasomes and mitochondrial import machines (18, 20, 21).

These broad roles of periodic biomechanical forces contrast sharply with little available knowledge on the response of protein structures to repetitive mechanical perturbation. Theoretical accounts on periodic biomechanics were reported as early as two decades ago (25). This work lay dormant for many years because of the absence of modern single-molecule measurements and computer simulations. More recently, several groups included periodic forces in numerical models of non-covalent receptor-ligand bonds and enzyme kinetics (26-29). They found optimal ranges of force frequency and amplitude that accelerate bond breakage or product formation. However, these kinetic models built on energetics determined in non-periodic experiments and did not include dynamic protein structures. Fundamental questions remain unanswered: How do protein structures respond to periodic forces of different frequencies and amplitudes? Do periodic forces lower the stability of proteins and/or shift unfolding pathways in their energy landscapes? Are these effects occurring on scales that are relevant in the context of biological machines?

Molecular dynamics simulations have proven essential tools in the study of the nano-mechanics of biomolecules and their interactions (15). Here, we use a coarse-grained approach that combines a mechanical pulling protocol, which is commonly used to probe protein nano-mechanics in simulations and experiments, with periodic forces. This approach allow us to gain first quantitative insights into the dynamic mechanical stability of single and multi-domain proteins. For the model system ubiquitin, we observed a complex response that included mechanical weakening of the protein, shifts in unfolding pathways and refolding.

Our observations are captured using energy landscape models and discussed in light of single-molecule measurements and physiologically relevant periodic forces.

## Results and Discussion

### Coarse-grained simulation approach to periodic forces

Proteins experience periodic mechanical forces in many physiological contexts (3, 9, 18-24). Here, we devised a computational approach to probe protein nano-mechanics in response to repetitive mechanical perturbations. To obtain complete pictures of dynamic protein stability, one needs to model repetitive forces with periods that are both much longer and shorter than mechanical protein relaxation times that occur on nanosecond time scales. Since it is challenging to access simulation times several orders of magnitude longer than nanoseconds in all-atom representations, particularly if many trajectories are required, we based our molecular dynamics approach on a coarse-grained Go-like model (30, 31). Coarse-grained models are widely used to study protein (un)folding on long time scales (32, 33) and geometry-based Go-like models have been shown to accurately reproduce experimental and atomistic computational studies of mechanical protein denaturation (34, 35). In our approach, we included a sinusoidal driving force of a wide range of frequencies and amplitudes in a constant velocity (CV) 'pulling' protocol, which is commonly used in single-molecule experiments and simulations (15). A first spring flanking the protein oscillates in z direction with amplitude $A$ (in units of distance, or $F_0$, if in units of force) and frequency $v$ ($\delta z = A\sin 2\pi v t$) while a second spring moves with constant velocity ($\delta z = Vt$, **Fig. 1A**). Using this coarse-grained methodology we were able to apply repetitive forces with periods almost two orders of magnitude longer than the relaxation times of single proteins (see below). The implementation of a CV protocol has two additional advantages. Firstly, we can compare our results to mechanical single-molecule experiments of CV protein unfolding combined with periodic forces. Furthermore, we adjusted the pulling velocity such that the number of oscillations per unit biomolecular length ranges from $2\text{nm}^{-1}$ to $20\mu\text{m}^{-1}$ and is thus in agreement to biological machines like the packaging motor of the bacteriophage Phi29 ($\approx 0.5\text{nm}^{-1}$) (9) or the bacterial ClpXP proteasome ($\approx 2\text{nm}^{-1}$) (20, 36).

### Periodic forces weaken ubiquitin

We choose a single ubiquitin domain (Ubq-1) as our first model system since there is a large body of experimental and theoretical data on ubiquitin's mechanical stability (37-42). We performed simulations on Ubq-1 with periodic oscillations of four different frequencies ($v=6.3\mu\text{s}^{-1}$, $63\mu\text{s}^{-1}$, $630\mu\text{s}^{-1}$ and $6.3\text{ns}^{-1}$) and 5 amplitudes (0.5, 1.0, 1.5, 2.5 and 3.75nm) applied to the C-terminus of Ubq-1. The frequencies were chosen such as to probe different regimes of the protein's dynamic response. Since the characteristic mechanical relaxation time of folded Ubq-1, $\tau_r$, is ≈3ns in our model (also see Materials and Methods), $v$=6.3ns$^{-1}$ corresponds to a high-frequency regime, $v$=630μs$^{-1}$ to an intermediate regime, and $v$=6.3 and 63μs$^{-1}$ to a low-frequency regime. **Fig. 1B** shows a force-displacement curve of Ubq-1 unfolded with periodic oscillations ($A$=2.5nm, $v$=63μs$^{-1}$, red trace), together with a curve for CV unfolding (black trace). Note that force traces reported here were averaged over the oscillation period. The most pronounced effect of periodic forces is a weakening of Ubq-1, i.e. reduced unfolding forces compared to CV simulations. A complete frequency- and amplitude-dependent analysis of the decrease in the most probable unfolding force, *δF*, is shown in **Fig. 2**. For reasons discussed below, we focused our systematic analysis on the main unfolding peak of Ubq-1, which occurs at extensions of ≈2-5nm and can be associated to a mechanical clamp formed between β-strands 1 and 5 (see below). From **Fig. 2A** it becomes clear that increased oscillation amplitudes result in decreased unfolding forces. For instance, in the low-frequency regime ($v$=63μs$^{-1}$), unfolding forces are only marginally lowered for $A$=0.5nm while for $A$=3.75nm the decrease in forces is 79.6 ± 5.8pN (or ≈46% of the CV unfolding force of 172.2 ± 1.5pN). We also observed that the impact of low-frequency modulations is significantly stronger than that for high frequencies. In case of $v$=6.3ns$^{-1}$, even amplitudes of 3.75nm do not lower unfolding forces significantly (**Fig. 2A**). Along these lines the results indicate the existence of an empirical frequency-dependent threshold value of amplitude below which the modulations have no effect on the unfolding force. For $v$=6.3 and 63μs$^{-1}$ the threshold is ≈1.0nm, whereas for $v$=630μs$^{-1}$ it shifts to ≈2.5nm. Intuitively, the frequency dependent modulation of Ubq-1 may be understood by noting that for high frequencies the changes in the oscillating force are too rapid to be followed by the protein that effectively feels only the constant (DC) component. Similar trends are observed for the second and third unfolding peak (data not shown). However, detailed analysis of these peaks is impeded because the unfolding pathways of Ubq-1 change in an amplitude and frequency dependent manner and because these unfolding intermediates are connected to the periodic force by a linker with non-linear elasticity (also see below).

**Analytical solution for protein unfolding in a periodically driven system**
To quantitatively understand the amplitude and frequency dependent stability of Ubq-1 we solved a kinetic model for protein unfolding in the presence of a sinusoidal force. The assumption of a sinusoidal force is not restrictive, since any periodic function can be reduced to a sum over sinusoidal components by Fourier analysis. Let us first consider an unbinding or unfolding process under the action of a DC force, $F$, modulated with a time-dependent (AC) force of amplitude $f_0$ and frequency $v$. The relevant timescales are the period of the AC force, $T_0 = 1/v$, the relaxation time of the protein, $\tau_r$, and the lifetime of the folded state, $\tau_F$. If $\tau_r$ is short in comparison to $T_0$, i.e. $\tau_r/T_0 \ll 1$, the reaction rate follows adiabatically the instantaneous potential and is given by (43, 44)

$$\kappa = \kappa(F)e^{\beta f_0 \sin 2\pi v t \delta x_t} \quad [1]$$

where $\beta = k_B T^{-1}$, $\delta x_t$ is the location of the transition state along the reaction coordinate and $\kappa(F) = \kappa_0 \exp(\beta F \delta x_t)$ is the unfolding rate in the DC case with $\kappa_0 = 1/\tau_F$. As long as $T_0$ is short in comparison to $\tau_F$, i.e. $T_0/\tau_F \ll 1$, the relevant quantity is not the instantaneous unfolding rate, $\kappa$, but rather the effective rate, $\kappa^*$, averaged over the period of the force

$$\kappa^* = \frac{\kappa(F)}{T_0}\int_0^{T_0} e^{\beta f_0 \sin 2\pi v t \delta x_t} dt = \kappa(F) I_0(f_0/f_\beta) \quad [2]$$

where $I_0$ is the modified Bessel function of the first kind and $f_\beta = (\delta x_t \beta)^{-1}$ is the characteristic thermal activation force. In the small noise intensity limit considered here, $f_0 > f_\beta$, and using the asymptotic form of the Bessel function for large arguments, $I_0(z) \approx \frac{1}{\sqrt{2\pi z}}e^z + ...$, we obtain

$\kappa^* = \kappa(F)\frac{1}{\sqrt{2\pi \beta f_0 \delta x_t}}e^{\beta f_0 \delta x_t}$. Thus the unfolding rate of the protein is enhanced with respect to the non-oscillating case by the factor which depends exponentially on the amplitude of the periodic force.

The problem becomes more complicated for higher frequencies, where adiabatic approximation no longer holds. Eq. **2** may then be generalized in the following form (45)

$$\kappa^*(F) = \kappa(F) I_0(\frac{f_0}{f_\beta}\frac{\tilde{\eta}(v)}{\delta x_t}), \quad [3]$$

where $\tilde{\eta}(v) = -\int_{-\infty}^{\infty} \frac{dx^0}{dt}e^{j2\pi v t}dt$, is the so-called logarithmic susceptibility with $x^0(t)$ being an optimal trajectory along the reaction coordinate. This optimal trajectory is a time-reversed trajectory from the transition state to the minimum of the potential well obtained by solving the equations of motion in the absence of both periodic modulation and random noise (45, 46). In the low-frequency limit ($v \to 0$), $\tilde{\eta}$ approaches $\delta x_t$ and Eq. **2** is recovered, while in the high-frequency regime ($v \to \infty$), $\tilde{\eta}$ approaches zero and the standard result corresponding to the escape over the unmodulated barrier is recovered. In this case the protein indeed is unable to trace the high-frequency force effectively responding to the DC component only.

In the context of mechanical unfolding by CV pulling, the force on the protein is continuously rising and $\kappa^*(F)$ increases with time. The quantity of interest is the most probable unfolding force, $\tilde{F}$, which may be obtained by considerations similar to those of Evans and Ritchie (44), this time, however, applied to the effective rate of Eq. **3**. This leads to $\tilde{F} = \tilde{F}_0 - \delta F$ where $\tilde{F}_0$ is the rupture force in the absence of modulation and

$$\delta F = f_\beta \log I_0 \left( \frac{f_0}{f_\beta} \frac{\tilde{\eta}(\nu)}{\delta x_t} \right) \quad [4]$$

is the force shift due to the periodic force. In the limit $f_0 \gg f_\beta$, we obtain $\delta F = \frac{1}{\delta x_t} f_0 \tilde{\eta}(\nu) - \frac{f_\beta}{2} \log 2\pi \frac{f_0}{f_\beta} \frac{\tilde{\eta}(\nu)}{\delta x_t} + \ldots$. Finally, in the adiabatic limit, $\nu \to 0$, $\delta F = f_0 - \frac{f_\beta}{2} \log 2\pi \frac{f_0}{f_\beta} + \ldots$, thus in this case the most probable rupture force is shifted with respect to the unmodulated case by $f_0$ up to the logarithmic correction. Contrastingly, $\delta F$ vanishes in the high-frequency limit.

Using Eq. **4** we can now analyze the measured force shifts (**Fig. 2B**). We only require an estimate of the thermal force scale, $f_\beta$=8pN (37), but not of the height of the energy barrier or the DC unfolding rate. In our simulations, we control the metric oscillation amplitude, $A$, and frequency, $\nu$, and the resulting amplitude of the modulating force is a function of those two parameters, $f_0(A,\nu)$, modulated by the elastic response of the protein. The force amplitude can be straightforwardly obtained by measuring the extension of the spring in the course of the simulation or in a test simulation at zero speed. For Ubq-1 the dependence $f_0(A)$ is linear with $f_0/A$=0.03, 0.31, 0.58 and 0.78 pN/nm for $\nu$=6.3$\mu$s$^{-1}$, 63$\mu$s$^{-1}$, 630$\mu$s$^{-1}$ and 6.3ns$^{-1}$ respectively (data not shown). Importantly, the only remaining unknown parameter in this model is the logarithmic susceptibility, which is determined when fitting with Eq. **4**. For $\nu$=6.3$\mu$s$^{-1}$, 63$\mu$s$^{-1}$, 630$\mu$s$^{-1}$ and 6.3ns$^{-1}$ we find $\eta(\nu)/\delta x_t$ = 0.98, 0.88, 0.23 and 0.02 respectively. Hence, as predicted, $\eta(\nu)$ approaches $\delta x_t$ in the adiabatic regime and zero in the high-frequency case. The quantitative agreement of this analytical one-barrier model and forces measured in simulations shows that the energy barrier confining the folded state is strongly lowered by the periodic force without indication of a shift in barrier position or more complex changes in the protein's energy landscape.

**Periodic forces shift unfolding pathways**

In addition to weakening we also observed marked changes in the unfolding pathways of Ubq-1. In CV simulations ($A$=0), three force peaks are observed (**Fig. 1B**, black trace) each of which corresponds to rupturing a group of secondary structure elements. We analyzed these unfolding events using scenario diagrams (**Fig. 3**), in which the last distance at which a native contact persisted is plotted against its contact order (the distance of the residues forming this contact as measured in primary sequence). The scenario diagram shows that the main force peak (occurring at ≈2-5nm extension) is associated with separating two pairs of parallel β-strands, namely S1 and S5 and S1 and S2. It has been previously shown that the S1-S5 structure acts as a mechanical clamp responsible for the protein's high mechanical resistance (39, 41). In a subsequent unfolding event, β-strands S3 and S5 are separated well after the other β-strands unfold. In this unfolding pathway (termed pathway 1, **Fig. 3A**) the unfolding sequence reads S1-S5, S1-S2, S3-S5, H1-H3, in agreement with existing experimental and theoretical studies on the mechanical denaturation of ubiquitin (37-42). As expected pathway 1 holds irrespective if the protein was pulled at the C- or N-terminus in CV-simulations at this pulling speed ($V$=3nm/μs). However, in the presence of periodic forces applied to the C-terminus we frequently noticed a second class of traces with only two peaks (**Fig. 1B**, red trace). Analysis using a scenario diagram showed that these traces originate from a new unfolding pathway (pathway 2, **Fig. 3B**). In pathway 2, β-strands S3 and S5 break early and cooperatively with the first event. Interestingly, this pathway is not populated at the lowest oscillation frequency ($v$=6.3μs$^{-1}$) which again indicates that protein stretching is adiabatic in this regime. For all other frequencies, increasing amplitudes result in pathway 2 being more populated (**Fig. 3F**).

The pathway shifts induced by periodic forces can be qualitatively described on an one-dimensional energy surface. In this simple model (**Fig. 4**, black trace), the native state and two unfolding intermediates are each confined by an energy barrier. In case of simulations with periodic forces, our data show that unfolding intermediate 1 (I1) is not populated and S3-S5 contacts break at the same time as S1-S5 and S1-S2 contacts. This suggests that in the presence of periodic perturbations the second energy barrier is not rate limiting (**Fig. 4**, blue trace) and we speculate that the protein unfolds 'downhill' to intermediate I2 after crossing the very first energy barrier. Such a 'downhill' model is also supported by above kinetic analysis of the very first force peak. Since a one-barrier model describes the force shift of this peak for all amplitudes (i.e. irrespective of the unfolding pathway) S3-S5 contacts likely never contribute to the first energy barrier.

One could now further speculate that the second energy barrier is strongly reduced by the periodic force because the protein is rapidly loaded with large transient pulling forces (**Fig. 2B**). To directly test this hypothesis we compared pathway 2 to unfolding pathways obtained when pulling at the C-terminus with faster velocities ($V$=30nm/μs) but no periodic force (**Fig. 3C**). A clear picture emerged and pathway 2 was observed in 74% (N=30) of the traces in line with above speculation. The remaining 26% of traces showed an unfolding pathway similar to pathway 1 (termed pathway 1') which is also detected when unfolding Ubq-1 from the N-terminus with high pulling velocities (see next section).

**Asymmetric protein response and refolding**

We also applied periodic forces to the N-terminus of Ubq-1 for an intermediate amplitude ($A$=2.5nm) and two frequencies ($v$=6.3$\mu s^{-1}$ and 63$\mu s^{-1}$). As expected for adiabatic unfolding at $v$=6.3$\mu s^{-1}$, pathway 1 was dominant by populating in 90% of the traces (pathway 2 in 10% of the traces, N=30). Thus the protein unfolded in the same manner as for both CV pulling and low-frequency forces applied to the C-terminus which indicates that the system is uniformly stretched and compressed. However, for $v$=63$\mu s^{-1}$ scenario diagrams showed an apparently different unfolding pathway (termed pathway 3) in 77% of the traces (pathway 2 in 23% of the traces, N=30). Here, contacts between β-strands S1 and S2 break at the very end of the unfolding process (**Fig. 3D**). The late rupture of S1-S2 contacts is in disagreement with the pathway observed for rapid unfolding by pulling at the N-terminus (100% pathway 1' at $V$=30nm/$\mu$s, N=30, **Fig. 3E**). In pathway 1', S1-S2 contacts break at extensions of ≈ 5nm in accord with their location close to the N-terminus (**Fig. 1A**). Since scenario diagrams only show the last time point at which a contact persisted we inspected the time evolution of S1-S2 contacts more closely. This analysis revealed that also in the case of a N-terminal periodic force S1-S2 contacts break at extensions of ≈5nm but then reform in a refolding event (**Fig. 5A**). This partial refolding of the N-terminal region of Ubq-1 was observed in all traces. Once this structure is formed it has a remarkable stability and persists until the protein is extended to ≈15nm. In apparent contrast refolding was not detected for other secondary structure elements (**Fig. 5B-D**).

There is a small but significant difference between pathway 1 and 1' which points towards pathway shifts induced by the N-terminal periodic force. In pathway 1' (and also 3), simultaneous rupture of contacts between α-helices H1 and H3 and contacts between β-strands S3 and S5 can be observed (**Fig. 3D and E**). Bonds between the helices H1 and H3 are apparently under large stress and speculatively the third energy barrier is lowered accordingly (**Fig. 4**).

Finally, marked difference between unfolding pathways 1' and 3 (observed with N-terminal perturbations) on one hand and pathway 2 (observed with C-terminal perturbations) on the other hand allows us to propose that Ubq-1's response to periodic forces is asymmetric. In case of the C-terminal periodic force, β-strands S3 and S5 located close to this terminus seem to be under large stress resulting in strong modulation of their energy barrier and early unfolding (see above). For N-terminal perturbations the β-strands S1 and S5 located close to this terminus are unfolded and refolded by the periodic force and the energy barrier for H1-H3 contacts is lowered. In these examples periodic forces with $v \geq 63 \mu s^{-1}$ effectively end up probing 'local' interactions in apparent contrast to adiabatic oscillations where stretch-compression cycles are more uniform. It is interesting to note that local interactions determine protein stability *in-vivo*, e.g. during protein unfolding at the proteasome or protein import into mitchondria (7).

**Periodic forces modulate the fingerprint of multi-domain proteins**

We also examined a multi-domain protein composed of three ubiquitin domains (Ubq-3). Mechanical responses of multi-domain proteins are more than serial combinations of single domains (47, 48) and have physiologically important mechanical signatures (49, 50). Here, we took advantage of multi-domain proteins to understand how linker elasticity influences the propagation of a periodic force in a biologically relevant mechanical system. **Fig. 1C** compares a CV unfolding trace of Ubq-3 with that obtained with periodic forces applied to the C-terminus. A very similar behavior as for Ubq-1 is observed. At $A$=2.5nm and $v$=6.3$\mu s^{-1}$ (63$\mu s^{-1}$) peak forces are significantly lowered by 64±6 (58±3), 41±5 (12±5) and 39±4 (5±4) pN for the domain that unfolds first, second and third respectively (N=24). While the force shifts of the first domain are considerable one observes much smaller and similar force shifts for the second and third domain. This can be explained if one considers that each denatured domain acts as a flexible, non-linear linker. Since the constant oscillation amplitude induces larger periodic forces in a shorter (stiffer) chain than in a longer (softer) chain the effect of periodic modulations becomes weaker with each unfolded domain. For this reason long unstructured domains, such as those found in some force bearing proteins (49, 50), should 'filter' periodic forces unless the protein is stretched to extreme lengths. Furthermore, the force fingerprint of a multi-domain protein is modified by periodic perturbations. In particular, one now observes an unfolding hierarchy where domains unfold at gradually increasing forces. Such a hierarchical pattern is not observed in CV simulations but governs the elasticity in some native multi-domain proteins, e.g. the distal tandem Immunoglobulin (Ig91) region of titin (47). We also analyzed the unfolding pathways of Ubq-3 and found significant pathway shifts for $v$=63$\mu s^{-1}$ but not $v$=6.3$\mu s^{-1}$. At $v$=63$\mu s^{-1}$ pathway 2 is populated with 92% and 75% for the domain that unfolded first and second (N=12). In both CV and periodic force simulations we were unable to clearly detect unfolding pathways in the domain that unfolded last.

**Correlation to mechanical single-molecule experiments**

In the past decade mechanical single-molecule experiments started to reveal the force-induced unfolding pathways and kinetics of water soluble and membrane proteins (51-53). Of particular interest here are modified atomic force microscopy (AFM) techniques developed to probe visco-elastic properties of single proteins (54-57). In these experiments, the AFM cantilever is sinusoidally oscillated resulting in a periodic force applied to the protein. Such experiments have not been conducted on ubiquitin, but we were nevertheless able to identify some of principles described above on other proteins. As predicted, Higgins *et al.* noticed a lowering of unfolding forces during the periodic force unfolding of an octamer of Ig91 domains with a frequency much lower than the relaxation time of the protein (55). The peak unfolding forces were independent of the number of unfolded domains which can be explained by considering that all analyzed domains were connected to the force probe through a flexible linker (see above). The experiments of Higgins *et al.* and of others also revealed that periodic forces change unfolding pathways of proteins like Ig91 or bacteriorhodopsin (54-57). In agreement with our observations on Ubq-3, pathway shifts were observed for all domains of multimeric proteins even if the unfolding forces of the domains were unaltered (54, 57). Our simulations thus capture many essential principles of experimentally determined response of proteins to repetitive forces.

**Conclusions**

Very little is known about how proteins respond to mechanical perturbations with repetitive components. Using molecular dynamics simulations we found that even a small sinusoidal force can trigger complex behaviors in a simple protein. In the model system ubiquitin, a periodic force weakened the protein's mechanical clamp, changed its unfolding pathways and allowed for transient refolding of single secondary structures. As intuitively expected these effects were amplitude- and frequency-dependent. Perturbations with very low frequencies ($<100\mu s^{-1}$) are of particular biological interest: Conformational changes in proteins typically occur on millisecond timescales and proteins are likely unable to generate forces faster than their relaxation times. We found that these low-frequency oscillations significantly lower the stability of ubiquitin (for amplitudes >1nm) but do not alter its unfolding pathways. In apparent contrast, higher frequency forces tilted the energy landscape differently resulting in shifted unfolding pathways. Although biological forces are most likely not strictly sinusoidal, we found that already this most simple function is capable of eliciting a complicated biomolecular response. Furthermore, through Fourier analysis, one should be able to reduce any time-dependent force to a sum over trigonometric components. Since many cellular forces have repetitive signatures we consider deeper insights into dynamic biomechanics an important starting point to understand protein folding and stability in cellular contexts. Periodic force simulations with the goal to better understand conformational changes in motor proteins are currently underway.

## Material and Methods

**Coarse-grained model**

For our geometry-based coarse-grained model we followed the Gō-like (30, 31) implementation of Cieplak and co-workers (58). Ubq-1 (PDB-ID 1ubq) is represented by a chain of $C^\alpha$ atoms tethered along the backbone by harmonic potentials with a minimum at 3.8Å. The effective interactions between the residues are split into native and non-native interactions by checking for native overlaps between the enlarged van der Waals surfaces of the residues (59). The amino acids, $i$ and $j$, that do overlap are endowed with the effective Lennard-Jones potential $V_{ij} = 4\varepsilon\left[\left(\frac{\sigma_{ij}}{r_{ij}}\right)^{12} - \left(\frac{\sigma_{ij}}{r_{ij}}\right)^{6}\right]$ with energy scale $\varepsilon$ and pair-by-pair distance $r_{ij}$. The length parameters, $\sigma_{ij}$, are chosen such that the potential minima correspond pair-by-pair to the experimental distances between the respective residues in the native state. Non-native contacts are represented by hardcore repulsion in order to prevent entanglements. Additionally, a correct chirality of the protein chain is imposed by the angle-dependent term in the Hamiltonian. This coarse-grained model was recently validated by comparing experimental unfolding forces and pathways to those obtained in simulations (34, 60). This comparison also allows to estimate $\varepsilon \approx 6.7$pNnm. Here, the overdamped motion of amino acids in solvent is mimicked using a standard Brownian dynamics algorithm (61). Both ends of the protein are attached to harmonic springs with spring constant $k$=80pN/nm (**Fig. 1A**). In CV simulations, the first spring (connecting points A and A') is fixed whereas the second spring (connecting points B and B') is pulled with a constant velocity ($V$=3 or 30nm/µs) along the initial end-to-end position vector. In periodic force CV simulations, point A is oscillated with amplitude $A$ and frequency $v$ (also see Results and Discussion).

**Estimating protein relaxation times**

We estimated the relaxation time of the protein towards equilibrium, $\tau_r$, in a simple test simulation. The native structure of a protein is perturbed by a small external force ($\approx$20pN) that stretches the bonds in the molecule but is too weak to disrupt native contacts. Then, by monitoring properties of the protein chain (end-to-end length and radius of gyration), we measure the time required to reach a new steady-state and use it as an order of magnitude estimate of $\tau_r$. For Ubq-1 $\tau_r \approx$3ns and for Ubq-3 $\tau_r \approx$27ns.

## Acknowledgments

We thank M. Cieplak, A. Kedrov and D.J. Müller for fruitful discussions. This research has been supported by grant N202 0852 33 of the Ministry of Science and Higher Education in Poland (to P.S.) and a fellowship of the European Molecular Biology Organization (to H.J.).

# Figures

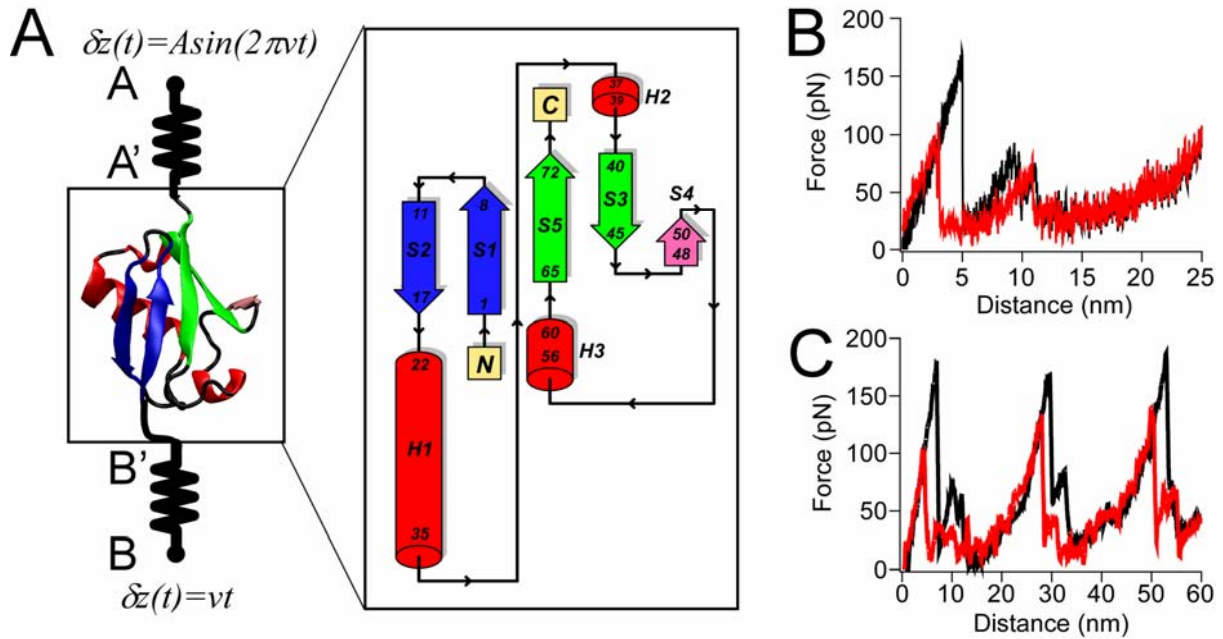

**Fig. 1.** Unfolding ubiquitin with periodic forces. (*A*) Schematics of periodic force simulations. Anchor A is oscillated with constant amplitude (*A*) and frequency (*v*) while anchor B is displaced with constant velocity (*V*). (Inset) Topology diagram of ubiquitin showing a mixed α-β-fold with α-helices H1 to H3 and β-strands S1 to S5. (*B* and *C*) F-D curves for Ubq-1 (Ubq-3) unfolding without (black traces) and with periodic forces (red traces, $A$=2.5nm, $v$=6.3µs$^{-1}$ for Ubq-1 and 63µs$^{-1}$ for Ubq-3). For visual clarity the curves are averaged over the period of force.

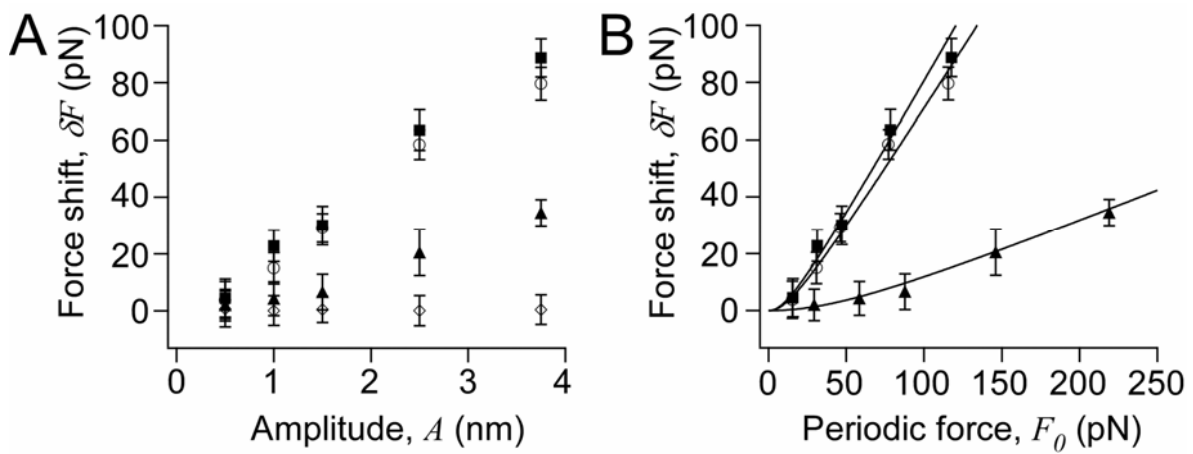

**Fig. 2.** Periodic forces weaken the mechanical clamp in Ubq-1. Change in peak unfolding force as a function of (*A*) oscillation amplitude and (*B*) force amplitude for $v$=6.3µs$^{-1}$ (squares), 63µs$^{-1}$ (circles), 630µs$^{-1}$ (triangles) and 6.3ns$^{-1}$ (diamonds). Low-frequency oscillations significantly lower the stability of the protein. In contrast, the protein does not respond to high-frequency oscillations resulting in marginally lowered unfolding forces and large force amplitudes. Solid lines in (*B*) are fits with Eq. **4**. Each datapoint is an average of 50 single molecule traces +/- SD.

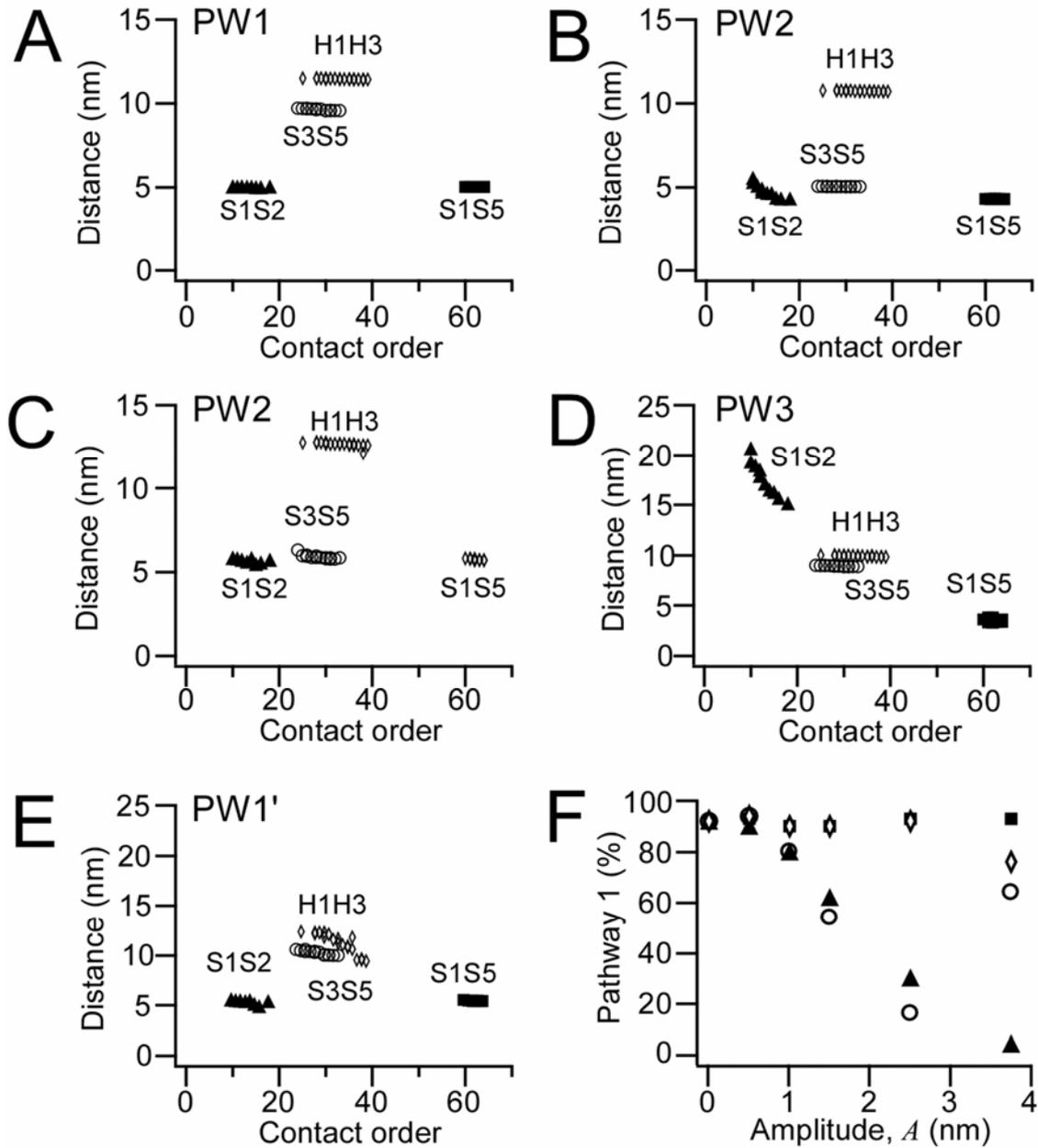

**Fig. 3.** Scenario diagrams reveal unfolding pathways of Ubq-1. Each symbol indicates the rupture of a native contact at a characteristic distance and contact order. (*A*) Pathway (PW) 1 is the dominant unfolding pathway measured when pulling at the C- or N-terminus without periodic forces ($V=3$nm/$\mu$s). (*B*) Pathway 2 is the dominant unfolding pathway in presence of periodic forces applied to the C-terminus ($A=2.5$nm, $v=630\mu s^{-1}$, $V=3$nm/$\mu$s). (*C*) Dominant unfolding pathway measured by pulling at the C-terminus with fast velocity ($V=30$nm/$\mu$s). (*D* and *E*) Unfolding pathway measured by pulling at the N-terminus with periodic forces (*D*, pathway 3, $A=2.5$nm, $v=630\mu s^{-1}$, $V=3$nm/$\mu$s) or fast velocity (*E*, pathway 1', $V=30$nm/$\mu$s). For visual clarity only contacts between major secondary structures (**Fig. 1***A*) excluding contacts between S2 and H1 or S3 and S4 are shown in *A* to *E*. (*F*) Change in population of pathway 1 of oscillation amplitude for $v=6.3\mu s^{-1}$ (squares), $63\mu s^{-1}$ (circles), $630\mu s^{-1}$ (triangles) and $6.3$ns$^{-1}$ (diamonds) with periodic forces at the C-terminus ($A=2.5$nm, $V=3$nm/$\mu$s).

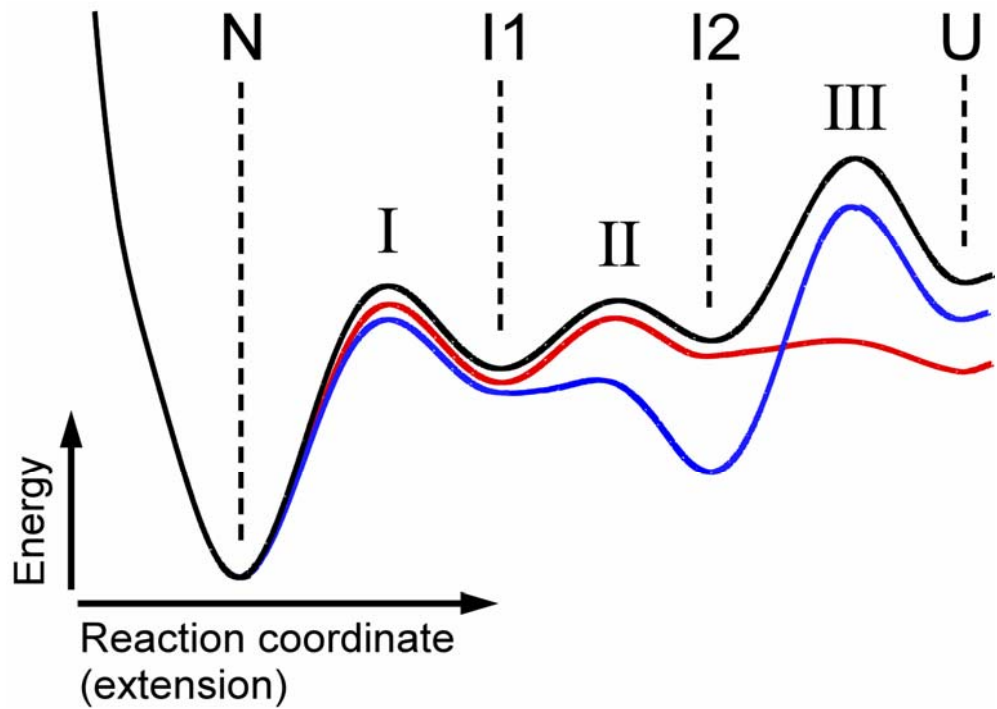

**Fig. 4.** Energy landscape for mechanical ubiquitin unfolding. Two unfolding intermediates (I1 and I2) separate the native state (N) from the unfolded state (U) on the energy surface (black line). Energy barrier I corresponds to contacts between two pairs of β-strands (S1 and S5 and S3 and S5) while energy barrier II and III confine contacts between β-strands S1 and S2 and α-helices H1 and H3, respectively. With a periodic force is applied to the C terminus I1 is not observed. We propose that large transient forces render barrier II not rate limiting allowing the protein to unfold 'downhill' to I2 (blue line). In apparent contrast, with a periodic force is applied to the N-terminus I2 is not observed and the protein unfolds directly to U (red line).

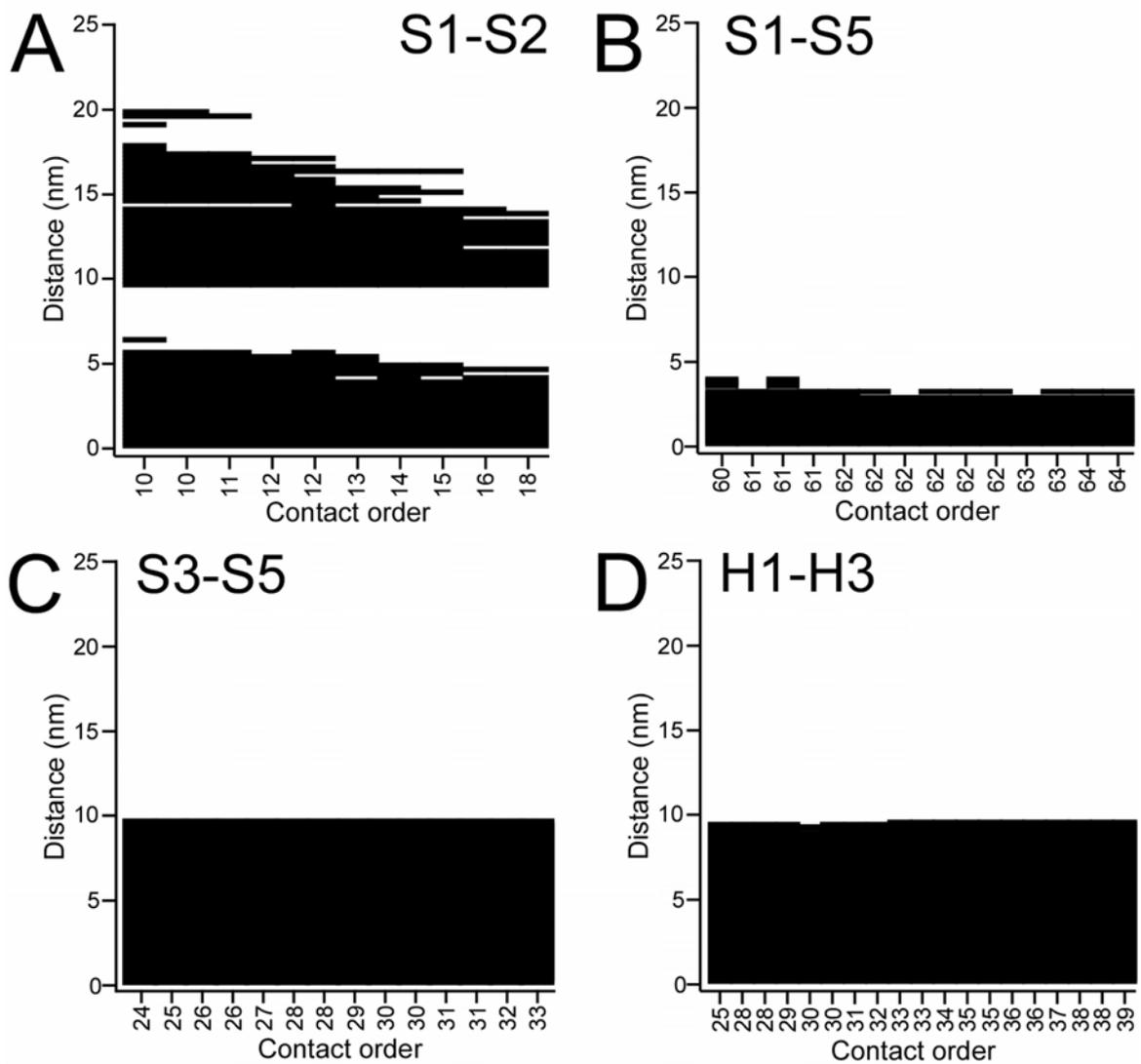

**Fig. 5.** Refolding analysis in a single Ubq-1 with a N-terminal periodic force. Contacts of four groups of secondary structures are shown sorted by contact order. Black squares indicate that a contact is present, while the absence of a square indicates that this contact is broken at a given displacement. (*A*) Transient refolding is observed for contacts between β-strands S1 and S2 after the initial rupture at ≈5nm distance. In contrast, contacts between other secondary structure elements (β-strands S1 and S5 (*B*), β-strands S3 and S5 (*C*) and α-helices H1 and H3 (*D*)) do not show refolding.